\documentclass[aps,pra,twocolumn,showpacs]{revtex4-1}

\usepackage{graphicx}
\usepackage{color}
\usepackage{amsmath}
\usepackage{epstopdf}
\graphicspath{{./figures/}}
\newcommand{\etal}{{\it et al.}~}

\begin{document}

\title{Inviscid diffusion of vorticity in low temperature superfluid helium}

\author{E. Rickinson$^1$}
\author{N. G. Parker$^1$}
\author{A. W. Baggaley$^1$}
\author{C. F. Barenghi$^1$}

\email{carlo.barenghi@newcastle.ac.uk}

\affiliation{$^1$Joint Quantum Centre (JQC) Durham--Newcastle, 
School of Mathematics, Statistics and Physics, Newcastle University, 
Newcastle upon Tyne, NE1 7RU, United Kingdom}

\date{\today}

\begin{abstract}
We numerically
study the spatial spreading of quantized vortex lines in low
temperature liquid helium. The vortex lines, initially concentrated 
in a small region, diffuse into the surrounding vortex-free helium,
a situation which is typical of many experiments. 
We find that this spreading, which occurs
in the absence of viscosity, emerges from the interactions between the
vortex lines, and can be interpreted as a diffusion process with
effective coefficient equal to approximately
$0.5 \kappa$ where $\kappa$ is the
quantum of circulation. 
\end{abstract}


\maketitle
    
\section{Introduction}

The work which we describe is driven by the 
comparison \cite{BSS-2014}
 between turbulence in ordinary fluids (classical turbulence)
and turbulence in superfluid helium (quantum turbulence). 
The main difference is the nature of the vorticity.
In ordinary fluids, vorticity is a continuous field, and 
vortices have arbitrary shape and strength. 
In superfluid helium, quantum mechanics constrains the vorticity to 
individual vortex lines of atomic thickness
(the vortex core radius is only $a_0 \approx 0.1~\rm nm$)
and fixed circulation $\kappa=h/m \approx 10^{-7}~\rm m^2/s$
(where $h$ is Planck's
constant and $m$ is the mass of one $^4$He atom).
In superfluid helium, turbulence thus
takes the form of a disordered tangle of interacting vortex lines.
Moreover, at temperatures below around $1~\rm K$, thermal excitations are negligible
and the vortex lines move in a perfect background fluid without viscosity.

Most experimental, theoretical and numerical studies 
have addressed quantum turbulence in its simplest form: 
statistically-steady, homogeneous and isotropic. These studies have revealed  
similarities and differences with respect to ordinary turbulence, in terms of energy 
spectra \cite{Salort-2010,BLR-2014,nocascade},
decay \cite{Walmsley-2008,Baggaley-2012}, 
intermittency \cite{Varga-2018,Rusaouen-2017,Boue-2013} and velocity 
statistics \cite{Paoletti-2008,Baggaley-2011-stats,LaMantia-2014-stats}.
Much less is known about turbulence which is inhomogeneous,
in particular 
turbulence which is initially confined in a small region of space and is
free to spread out.  A better understanding 
of this diffusion problem would help to interpret many helium experiments 
in which ultrasound \cite{Schwarz-1981}, 
oscillating spheres \cite{HanninenSchoepe-2008},
wires, grids \cite{Jackson-2017}
and forks \cite{Schmoranzer-2016,Bradley-2016}
create quantum turbulence in helium at rest, and from which the
turbulence may spread and fill the experimental cell. 
Particularly important, as already remarked, is the low temperature 
limit, in which the normal fluid is negligible and the dynamics of the
vortex lines is simpler, at least in principle. 

\begin{figure*}[ht]
\centering
\fbox{\includegraphics[width=0.33\linewidth]{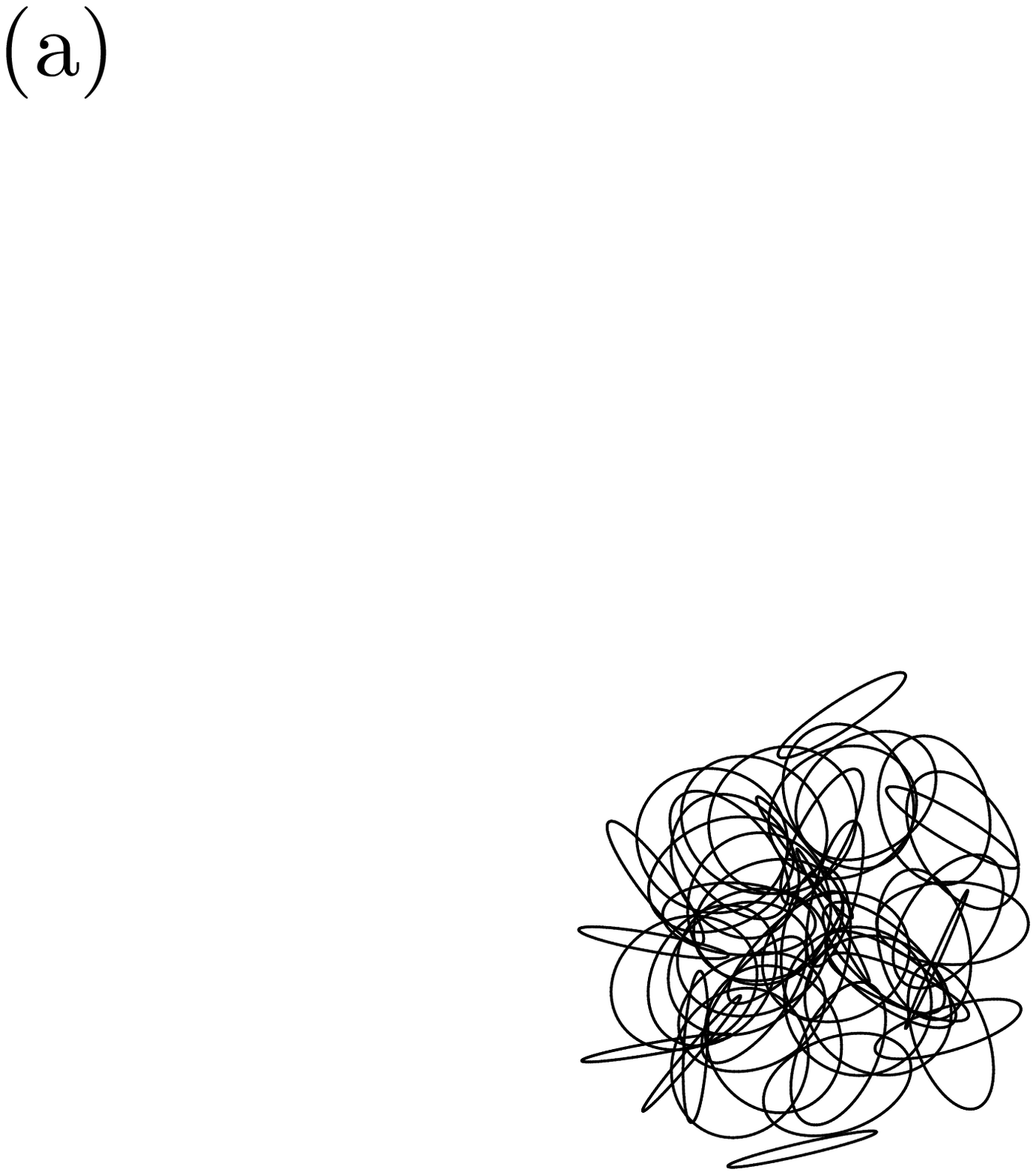}}
\fbox{\includegraphics[width=0.33\linewidth]{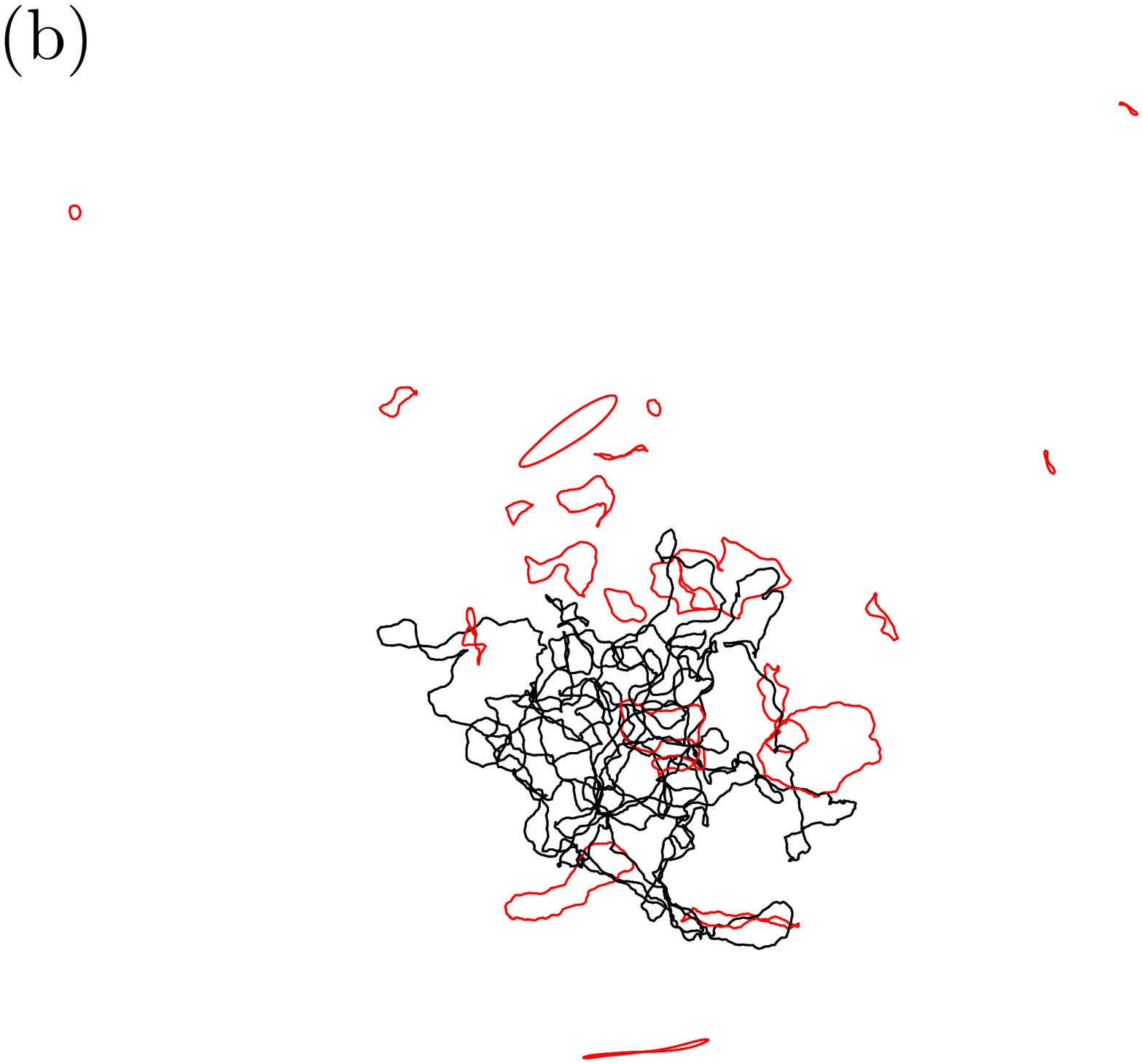}}\\
\fbox{\includegraphics[width=0.33\linewidth]{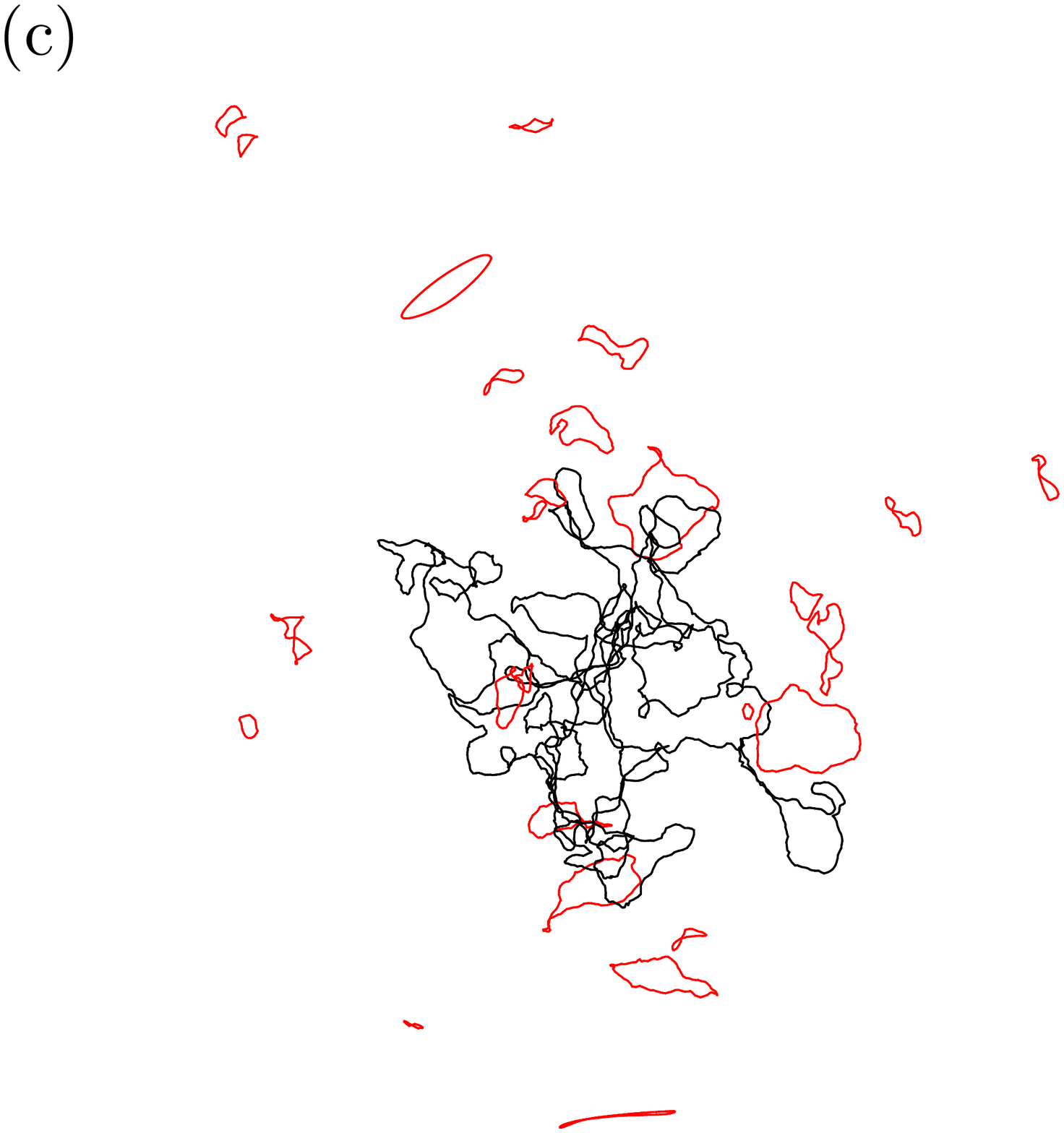}}
\fbox{\includegraphics[width=0.33\linewidth]{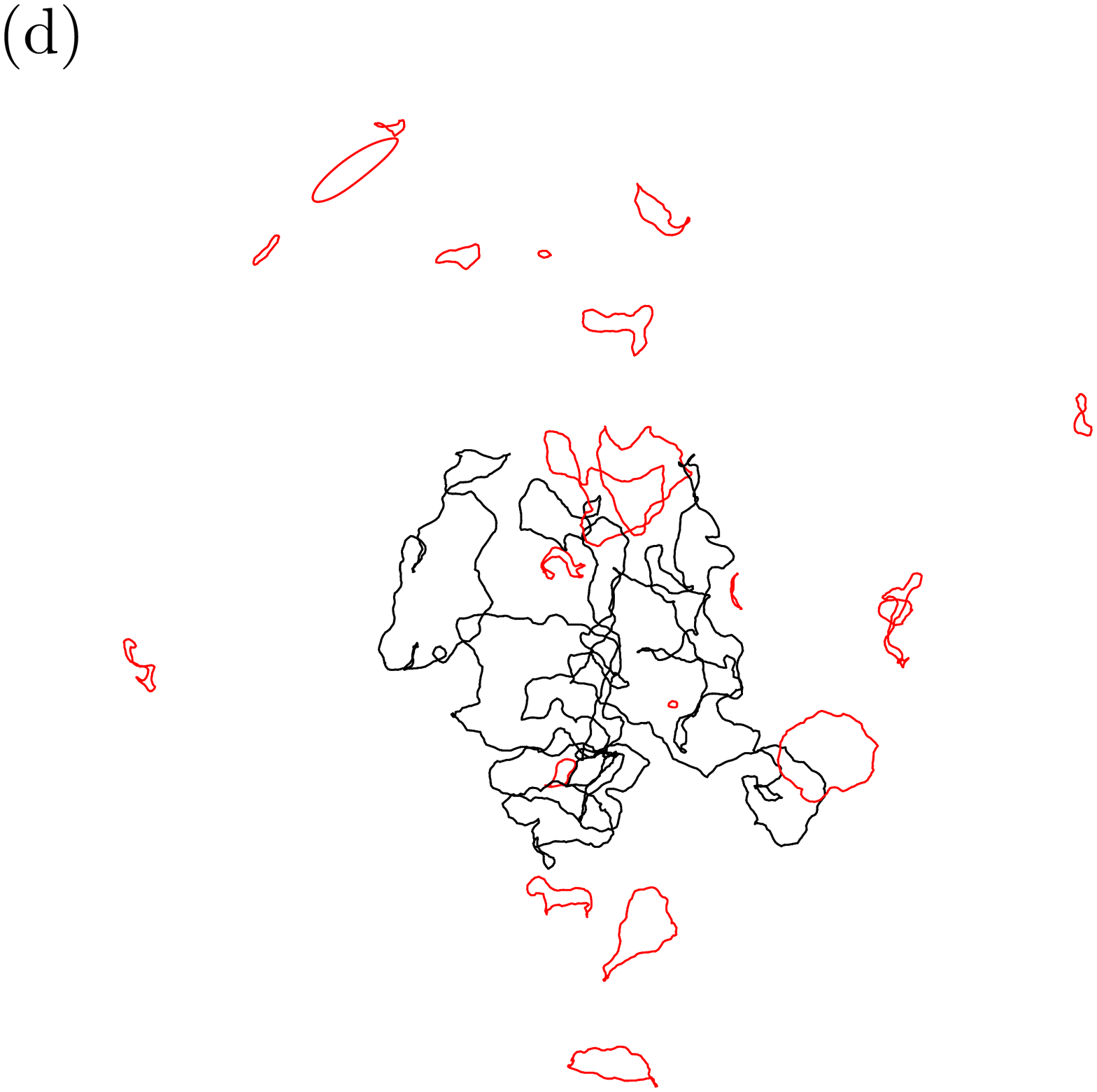}}
\caption{Typical time evolution of the vortex tangle:
(a): $t=0~\rm s$, (b): $t=100~\rm s$, (c): $t=200~\rm s$, and
(d): $t=300~\rm s$. All panels show the region 
$-2.5~\textrm{cm}\leq x,y \leq 2.5~\textrm{cm}$ projected onto the $z=0$ plane; The computational domain is infinite (no boundaries). Note how the vortex
tangle spreads out. Vortex loops containing fewer than 200 points are 
shown in red, with the remaining vortices shown in black.} 
\label{fig1}
\end{figure*}

In this article we report the results of numerical simulations of
the diffusion of a turbulent tangle of vortex lines which is 
initially localised in a region at the centre of the computational domain. 
In the related context of two-dimensional (2D) trapped Bose-Einstein 
condensates, using the Gross-Pitaevskii equation (GPE) model,
we have found \cite{Rickinson-2018} that an initial vortex cluster
diffuses via two distinct mechanisms: the evaporation of small 
vortex-antivortex pairs (vortex dipoles)
 which quickly leave the vortex cluster, 
and the slower spread of the cluster itself. The latter effect
is like a classical diffusion process with
an effective viscosity $\nu^{\prime} \approx \kappa$.

The natural question which we address here is whether this
effect holds true in three dimensions (3D) in the context of 
superfluid helium. The extra dimension introduces
effects which are absent in 2D, such as vortex reconnections and 
Kelvin waves. A pioneering numerical study by
Tsubota \etal \cite{Tsubota-2003} 
of the 3D diffusion of a vortex tangle reported a value of
$\nu^{\prime}$ smaller than what we found in 2D, albeit using 
a different approach and different initial conditions.  
These differences add further
motivations to revisit the 3D diffusion problem.

\section{Methods}

\subsection{Evolution of the vortex tangle}

To numerically simulate the evolution of a tangle of
quantised vortex lines,
we use the Vortex Filament Method (VFM) of
Schwarz \cite{Schwarz1989}, a more realistic model of turbulent He~II
than the GPE (which gives a good
quantitative description of low-temperature
Bose-Einstein condensates \cite{book}).  The VFM describes vortex lines
as space curves ${\bf s}(\xi,t)$ (where $t$ is time and $\xi$
is arclength) which move according to the Biot-Savart law:

\begin{equation}
\frac{d{\bf s}}{dt}=
-\frac{\kappa}{4\pi}\oint \frac{({\bf s-r}) \times {\bf dr}}{\left|{\bf s-r}\right|^3},
\label{eq:BS}
\end{equation}

\noindent
where the line integral extends over all vortex lines.
Since there are no boundaries, all vortex lines
form closed loops.  Our VFM \cite{Baggaley-2011-stats} uses 
a Lagrangian discretization along the vortex lines, 
with discretization points continuously added or removed to 
maintain the chosen spatial resolution of $\delta = 0.015~\rm cm$.
Vortex loops consisting of 
less than 5 discretization points are removed, modelling the effects of
the small residual friction which is present even for $T<1~\rm K$.
The Biot-Savart integral is de-singularised
in a standard way \cite{Schwarz1989} based on the vortex core cutoff $a_0$.
The procedure for vortex reconnections is implemented algorithmically
\cite{Schwarz1989,Baggaley2012b}. The resulting system of
differential equations is integrated in time using a $3^{\textrm{rd}}$ order Adams-Bashforth scheme with a timestep of $5\times10^{-3}$s.

The typical initial condition, shown in panel (a) of Fig.~\ref{fig1}, 
consists of a set of randomly oriented vortex loops with radius $0.24~\rm cm$, 
randomly and independently translated in the $x$, $y$, and $z$ directions
according to a normal distribution with 
standard deviation $1~\rm cm$. 
To explore the effect of changing the initial
vortex line density we perform two sets of 
simulations, one initialised with 50 vortex loops, leading to an initial 
vortex line density (vortex length per unit volume) of $L\sim70~\textrm{cm}^{-2}$ at the centre
of the infinite computational domain, and one using 100 vortex loops, 
with an initial vortex line density of 
$L\sim140~\textrm{cm}^{-2}$.

\subsection{Determining the effective diffusion}

We estimate the effective diffusion 
of the vortex tangle using two different techniques. 
The first technique follows the work of
Tsubota \etal \cite{Tsubota-2003}, who determined 
$\nu^{\prime}$ using the following
modified Vinen equation for a space-dependent vortex line density
$L({\bf x},t)$ :

\begin{equation}\label{eq:Vinen}
\dfrac{\partial L}{\partial t}=-\frac{\kappa}{2\pi}\chi_2 L^2 
+\nu^{\prime} \nabla^2 L.
\end{equation}

\noindent
The original Vinen equation \cite{Vinen1957} (see appendix) balances a generation
term (proportional to the driving counterflow velocity and $L^{3/2}$)
against a decaying term (proportional to $L^2$). 
Equation~\eqref{eq:Vinen}  contains the same
decaying term of the original Vinen equation (proportional to $L^2$), but
lacks the
generation term (because it is concerned with decay at zero temperature), and
postulated the existence of a diffusion
process represented by the new term $\nu^{\prime} \nabla^2 L$; 
this new term turns the original Vinen equation into a parabolic 
partial differential equation. 
In writing Eq.~\eqref{eq:Vinen}, Tsubota \etal 
assumed that $\nu^{\prime}$ depends on the 
temperature but is independent of the vortex line density.
To determine $\nu^{\prime}$, they fitted
the computed coarse-grained vortex line density to the numerical solution of
Eq.~\eqref{eq:Vinen}. The fit, however, requires knowledge of Vinen's parameter
$\chi_2$. Tsubota \etal estimated \cite{tsubota2000} that $\chi_2 \approx 0.3$
from separate numerical simulations at $T=0$ performed using the local 
induction approximation to the exact Biot-Savart law (Eq.~\eqref{eq:BS}).
Physically, in this zero temperature limit, $\chi_2$ models 
a sink of vortex lines due to dissipation 
of kinetic energy through both vortex reconnections and phonon emission 
(induced by high frequency Kelvin waves \cite{Baggaley2014}).
 
It has been noted that $\chi_2$ depends on the local vortex line density at finite temperatures \cite{Gao-2018}. We independently estimate values of $\chi_2$ in the zero-temperature limit, using the full Biot-Savart law of Eq. (\ref{eq:BS}), for $13$ values of $L$, finding $\chi_2(L)\approx0.07L^{0.4}$; this is detailed further in the appendix.
Assuming spherical symmetry, we then estimate $L(r)$ (where $r$ is the radial distance from the centre) by integrating over the vortex lines within 
spherical shells, subdividing the line segments for a more accurate 
measurement when they cross between shells, and dividing by the volume 
of these shells. We numerically solve the modified Vinen equation in a radially symmetric 
coordinate system, using $4^{\textrm{th}}$-order finite difference methods 
for spatial derivatives, and a $3^{\textrm{rd}}$-order 
Adams-Bashforth time integration scheme with 
timestep $\Delta t=10^{-2}\rm s$. 
We use a reflective boundary condition to enforce
$dL/dr=0$ at $r=0$, impose $L=0$ at $r=10~\rm cm$ 
(far from the region of interest), 
and use the initial vortex line density as a function of $r$ estimated 
from our VFM simulations as the initial condition. The local value of $\chi_2$ is taken to be $\chi_2(L)=0.07L^{0.4}$ based on our above estimates.

The second technique is based on considering the deviation in the trajectories of diffusing tracers of the flow \cite{Sikora-2017}, which in our context is provided by the individual vortex discretization points modelled by the VFM.  We know that the diffusion constant $\nu$ of
a scalar field $F(x,y,z,t)$ which satisfies
the diffusion equation
\begin{equation}
\frac{\partial F}{\partial t}=\nu \nabla^2 F,
\label{eq:diff}
\end{equation}
\noindent
is related to the root-mean-square (rms) deviation $d_{\rm rms}(t)$ by
\begin{equation}
\nu=\frac{d_{\rm rms}^2(t)}{4 t}.
\label{eq:conversion}
\end{equation}
We define the rms deviation of our $N_0(t)$ vortex discretization points 
from their initial positions as
\begin{equation}
\label{drms}
d_{\text{rms}}(t)=
\sqrt{\frac{1}{N_0(t)}  \sum^{N_0(t)}_{i=1} 
\left( \Delta x_i^2(t)+\Delta y_i^2(t)+ \Delta z_i^2(t) \right) }
\end{equation}
\noindent
where $\Delta x_i(t)=x_i(t)-x_i(0)$, $\Delta y_i(t)=y_i(t)-y_i(0)$, and
$\Delta z_i(t)=z_i(t)-z_i(0)$. 
\noindent
Using Eq.~(\ref{eq:conversion}) we can define an effective 3D diffusion
coefficient $\nu^{\prime}$ representing  the spatial spreading of the vortex
cluster.

Because discretization points along vortex lines are continually removed 
and added to maintain the spatial resolution, care must be taken 
in establishing their trajectories. In most cases there is a direct link 
between a point at a given time and the same point at the previous time 
(the ancestor). If a point is newly inserted, and thus lacks  an ancestor 
at the previous time,  we consider the ancestors of the discretization 
points on either side of the newly inserted point, and store both these 
values as ancestors of the new point, using the average initial position 
of these ancestors as the initial position of the new point. 
This process is iterative as ancestors are concatenated in successive 
time steps. This second technique generalizes our 
previous 2D work \cite{Rickinson-2018} to 3D.

\section{Results}

The typical evolution of the initial vortex rings into a turbulent
vortex cluster is shown in Fig.~\ref{fig1}. 
Before proceeding with the calculation of $\nu^{\prime}$, a
natural question arises: what is the character of this turbulence?
Usually the answer is given in terms of the energy spectrum, but in this
case the turbulence is neither steady nor homogeneous, and the interpretation
of the spectrum would be difficult. We proceed differently and
calculate the transverse velocity correlation function
$f_{\perp}(r,t)=\langle v_{\perp}({\bf x},t)v_{\perp}({\bf x}+r\hat{\bf e}_{\perp},t)\rangle/\langle v_{\perp}({\bf x},t)^2\rangle$, 
and find that it rapidly decreases with distance, meaning
that the turbulent velocity field is essentially random; at $t=0$,
we find that $f_{\perp}(\ell/2,0)\approx0.27$ only, 
where $\ell$ is the inter-vortex spacing,
indicative of the Vinen (ultra-quantum) regime of quantum 
turbulence \cite{Walmsley-2008}, 
characterized by the absence
of an energy cascade \cite{nocascade}. Similar turbulence and 
correlation functions have been
predicted in trapped atomic Bose-Einstein condensates \cite{Cidrim2017}.

We now consider how the turbulence spreads in space.
The initial vortex rings interact, become
distorted, and undergo vortex reconnections,
generating small vortex loops; if these small loops are
in the outer part of the cluster and are oriented outwards,
they quickly leave the cluster, as seen in the figure. 
This ``vortex evaporation" \cite{BS-2002} has been noticed in
experiments \cite{Fisher-2001} and reported
in other 2D and 3D numerical simulations \cite{Rickinson-2018,Stagg-2017}.
Here we concentrate on the slower spread of the main vortex
cluster. 

We first follow the approach of Tsubota \etal \cite{Tsubota-2003}, 
and seek the solution of Eq.~\eqref{eq:Vinen}, estimating $\nu^{\prime}$ 
by minimising the sum of 
square errors between the vortex line density estimated from the VFM 
simulations and the numerical solution of Eq.~\eqref{eq:Vinen}.
We find that, using this method,
our estimate of $\nu'$ is very sensitive to the initial vortex line 
density, and possesses considerable uncertainty.
Taking $\chi_2=0.07L^{0.4}$ gives $\nu'/\kappa=0.28\pm0.11$ for
the high density simulations, and $\nu'/\kappa=0.33\pm0.20$ 
for the low density ones.

\begin{figure}[h]
\includegraphics[width=1\columnwidth]{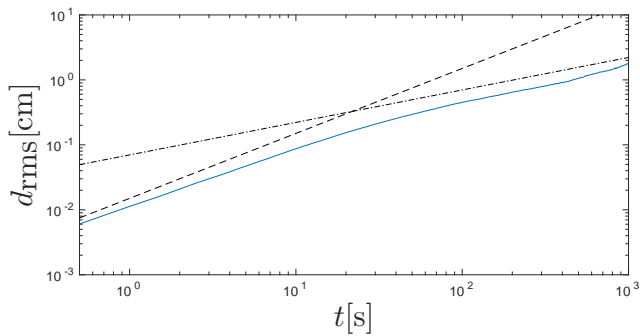}
\caption{Typical temporal dependence of the rms deviation of tracers from
their initial position, $d_{\rm rms}$ (in cm) vs. $t$ (in s). 
Notice the transition from ballistic ($d_{\rm rms} \sim t$) 
to diffusive ($d_{\rm rms} \sim t^{1/2}$) regimes, indicated 
by the dashed and dot-dashed lines respectively.
}
\label{fig2}
\end{figure}

We now turn to the second approach, in which we
infer the diffusion coefficient from individual trajectories of 
diffusing tracers from time averaged deviations \cite{Sikora-2017} 
defined in Eq.~(\ref{drms}).  The typical temporal behaviour of the rms deviation of 
tracers from their initial positions, $d_{\rm rms}$ vs. $t$, averaged over $10$ simulations, is shown in
Fig.~\ref{fig2}.
The figure shows that the initially ballistic  
regime, $d_{\rm rms} \sim t$, is followed by a $d_{\rm rms} \sim t^{1/2}$
diffusive regime. 

\begin{figure}[ht]
\includegraphics[width=1\columnwidth]{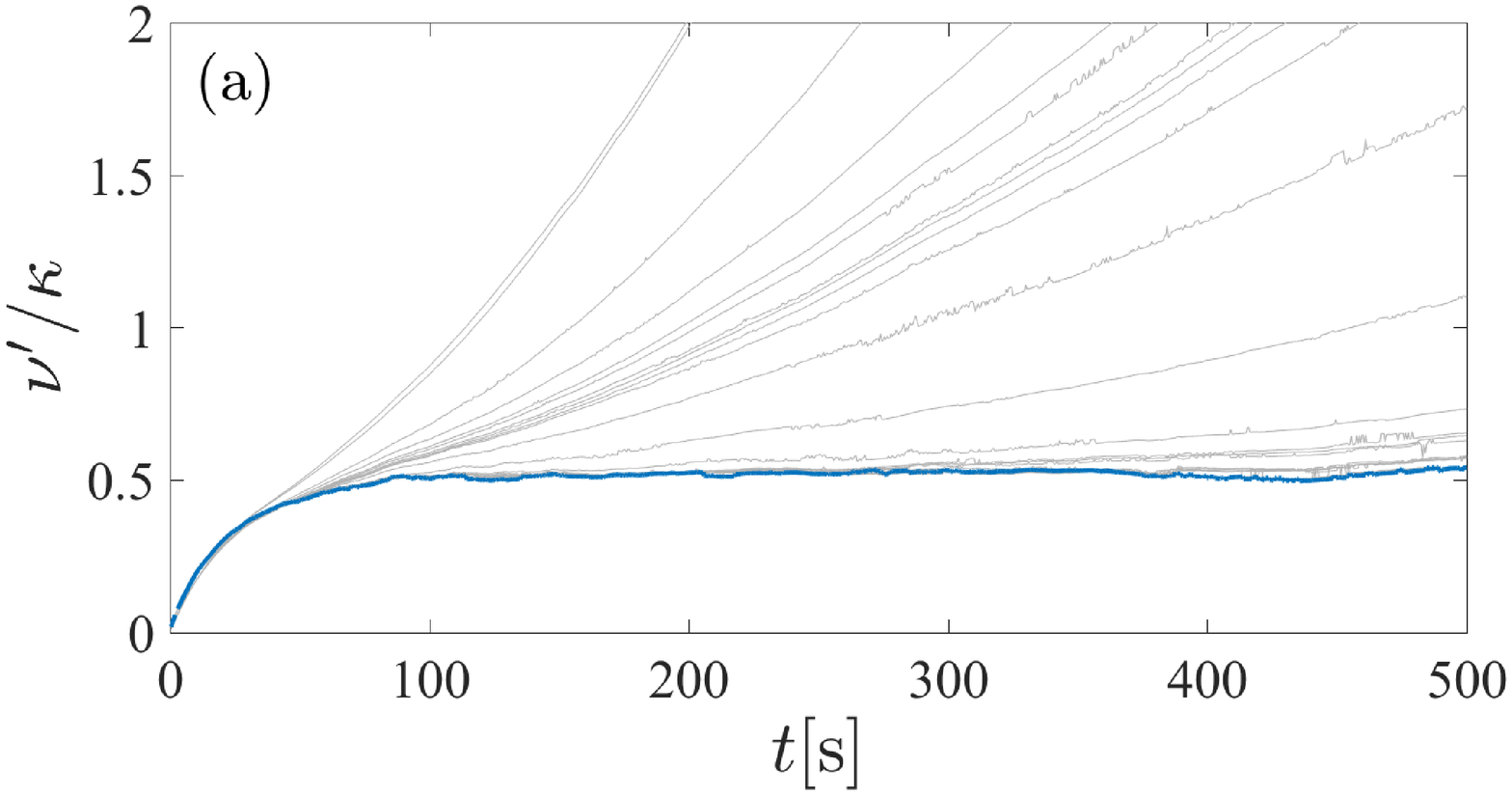}
\includegraphics[width=1\columnwidth]{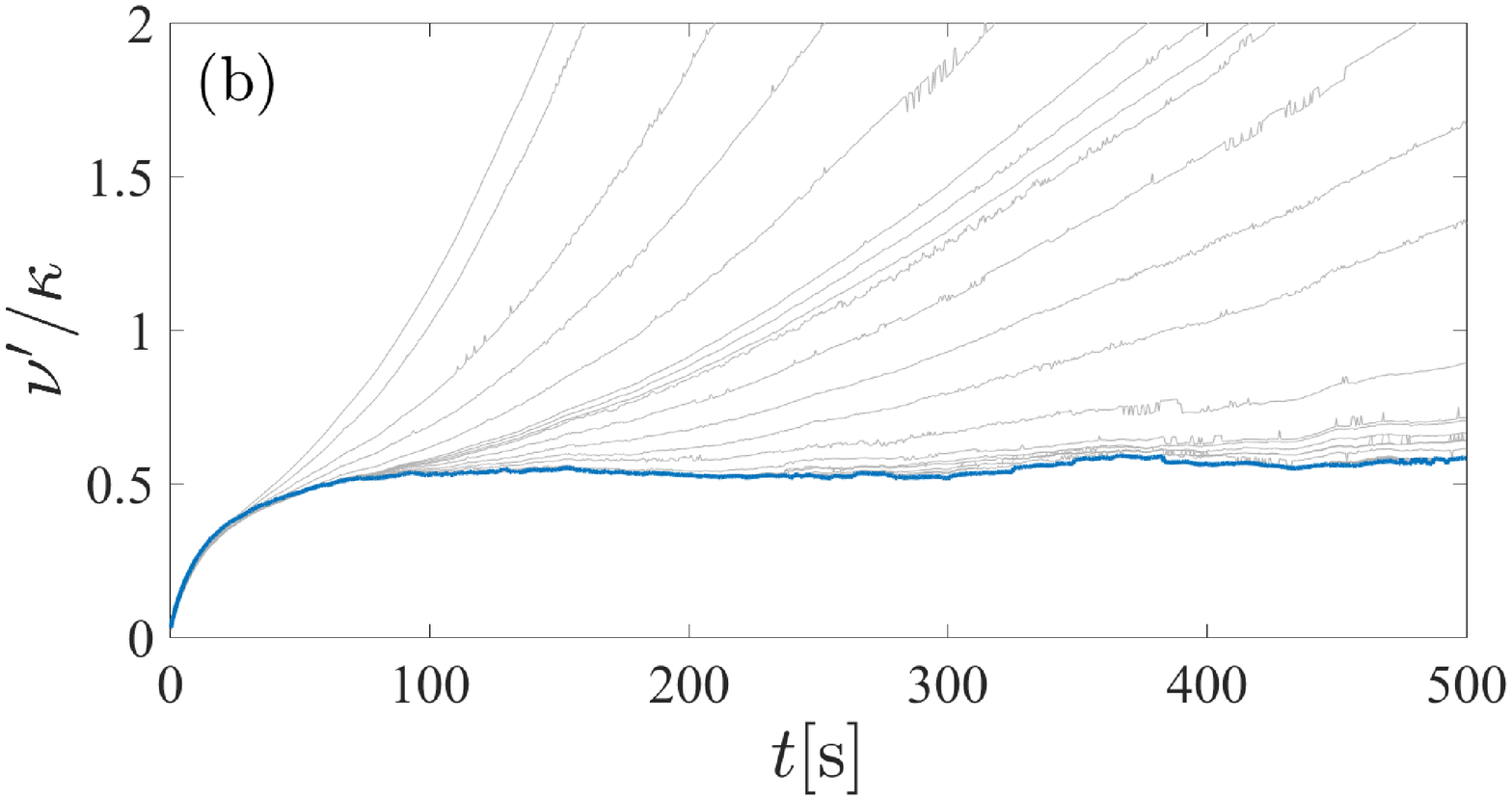}
\caption{Effective diffusion in units of the quantum of circulation, 
$\nu'/\kappa$, vs. time $t$ (in s) (solid blue line) for low density simulations (a) and high density simulations (b). Grey lines show 
the values of $\nu^{\prime}/\kappa$ found as the minimum size of loops 
included in our calculation of $d_{\textrm{rms}}$ is increased, 
from $0$ points (grey line which attains maximum value earliest) 
to $200$ points, for which  the value of $\nu^{\prime}/\kappa$ 
found has converged.}
\label{effectiveviscosity}
\end{figure}

\noindent
The effective diffusion
coefficient $\nu^{\prime}$, obtained from Eq.~(\ref{eq:conversion}), is plotted as a function of time $t$ in Fig.~\ref{effectiveviscosity}(a) 
(low density simulations) and
Fig.~\ref{effectiveviscosity}(b) (high density simulations) (solid blue line). 
It is apparent
that the effective diffusion settles down to the value
$\nu^{\prime}/\kappa \approx 0.5$ in both simulation sets. More precisely, we obtain $\nu^{\prime}/\kappa=0.526\pm0.064$ for low
vortex line density and $\nu^{\prime}/\kappa=0.530\pm0.065$ for high vortex
line density).

It is important to appreciate that, 
unlike the approach of Ref.~\cite{Tsubota-2003}, when we compute
$\nu^{\prime}$ via $d_{\text{rms}}$, we do not include
the tracers which belong to evaporating vortex loops  
as they move ballistically. The evaporating vortex loops, in fact,
do not interact strongly with the other vortices of the
cluster any longer, but move away with approximately constant
speed determined by their average curvature.
We observed an analogous effect in our previous
2D simulations \cite{Rickinson-2018}, where  vortex dipoles
(the 2D analog of 3D vortex loops) 
ballistically evaporate from the vortex cluster. In this previous 
work we used a numerical procedure to identify and remove 
these fast moving dipoles from the calculation of $\nu^{\prime}$.
Generalizing this 2D procedure to 3D, our analysis
neglects fast evaporating vortex loops
if they contain less than a certain critical number of 
discretization points $N_c$; 
in this way, we effectively set a minimum size for a vortex loop
to be included in the calculation of $d_{\rm rms}$ and $\nu^{\prime}$. The
critical number $N_c$ is empirically determined.
In Fig. \ref{effectiveviscosity} the grey lines show the values 
of $\nu^{\prime}/\kappa$ 
found at increasing $N_c$, from $N_c=0$ (grey line which attains maximum 
value earliest) to $N_c=200$ (solid blue line); note that for $N_c=200$
the value of $\nu^{\prime}/\kappa$ has converged. 
This distinction between evaporating loops and the remaining vortex
cluster is highlighted in Fig.~\ref{fig1}, where vortex loops containing 
fewer than 200 discretization points (hence small compared to other loops) are shown in red, while the remaining vortex lines are shown in black.

\begin{figure}[h]
\centering
\fbox{\includegraphics[width=0.4\linewidth]{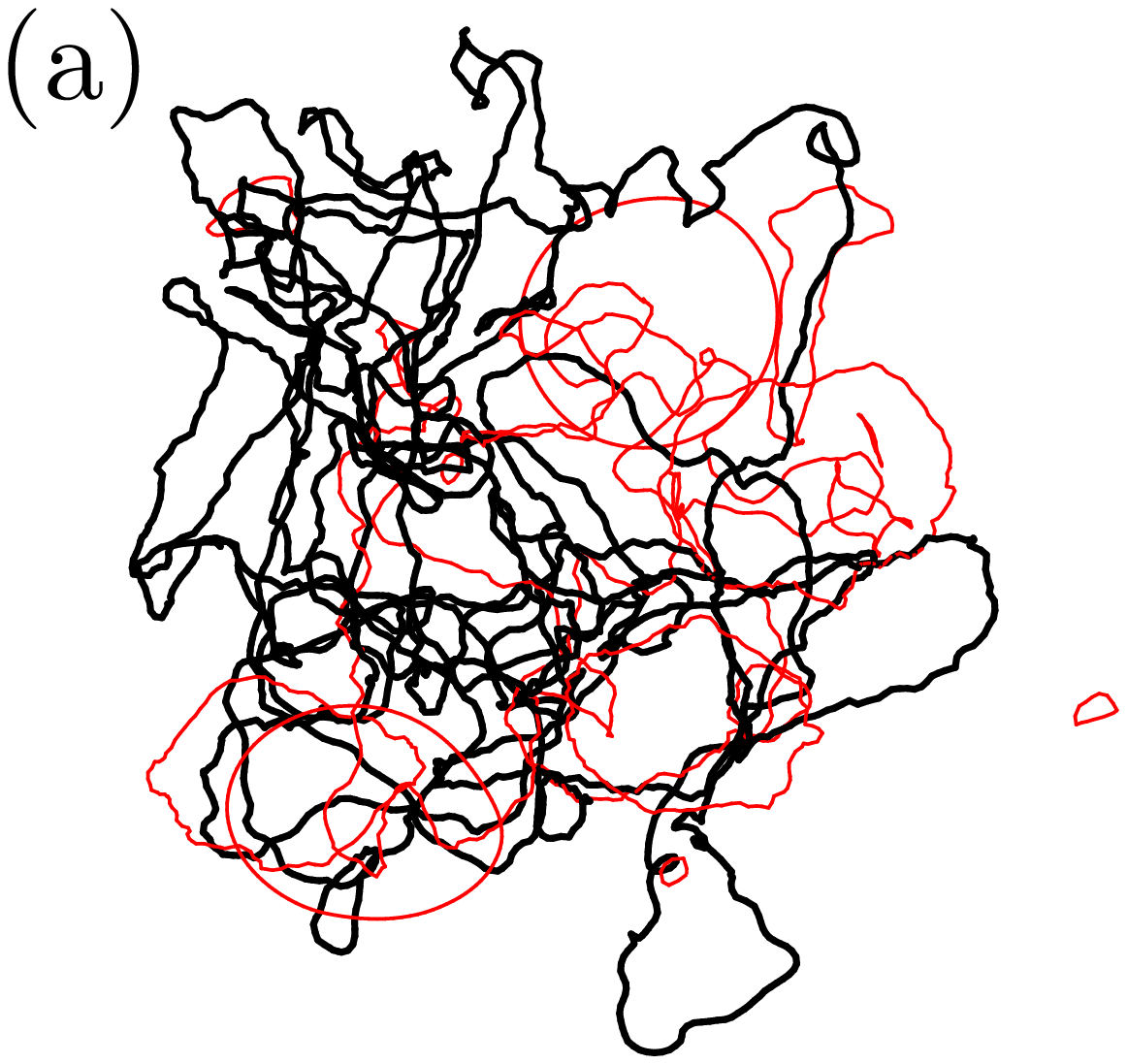}}\hspace{4pt}\fbox{\includegraphics[width=0.4\linewidth]{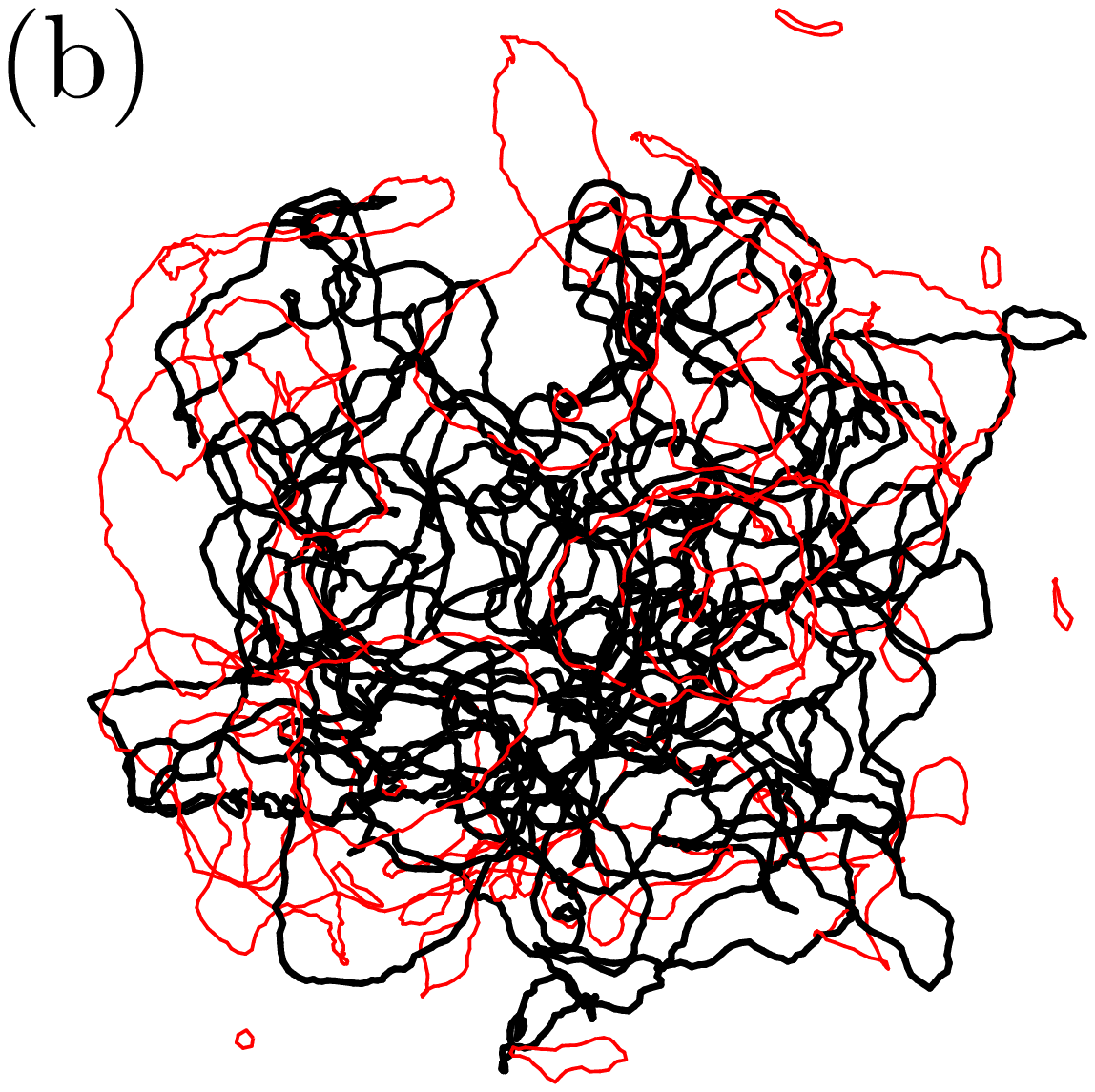}}
\caption{Typical large knotted vortex structures (black lines) from simulations at (a) lower density ($L(|\mathbf{x}|=0,t=0) \sim 70 {\rm cm}^{-2}$) and (b) higher density ($L(0,0) \sim 140 {\rm cm}^{-2}$), at $t=50$s .  The other vortex lines are shown as red curves.} 
\label{fig:knots}
\end{figure}

\begin{figure}[h]
\centering
\includegraphics[width=0.5\linewidth]{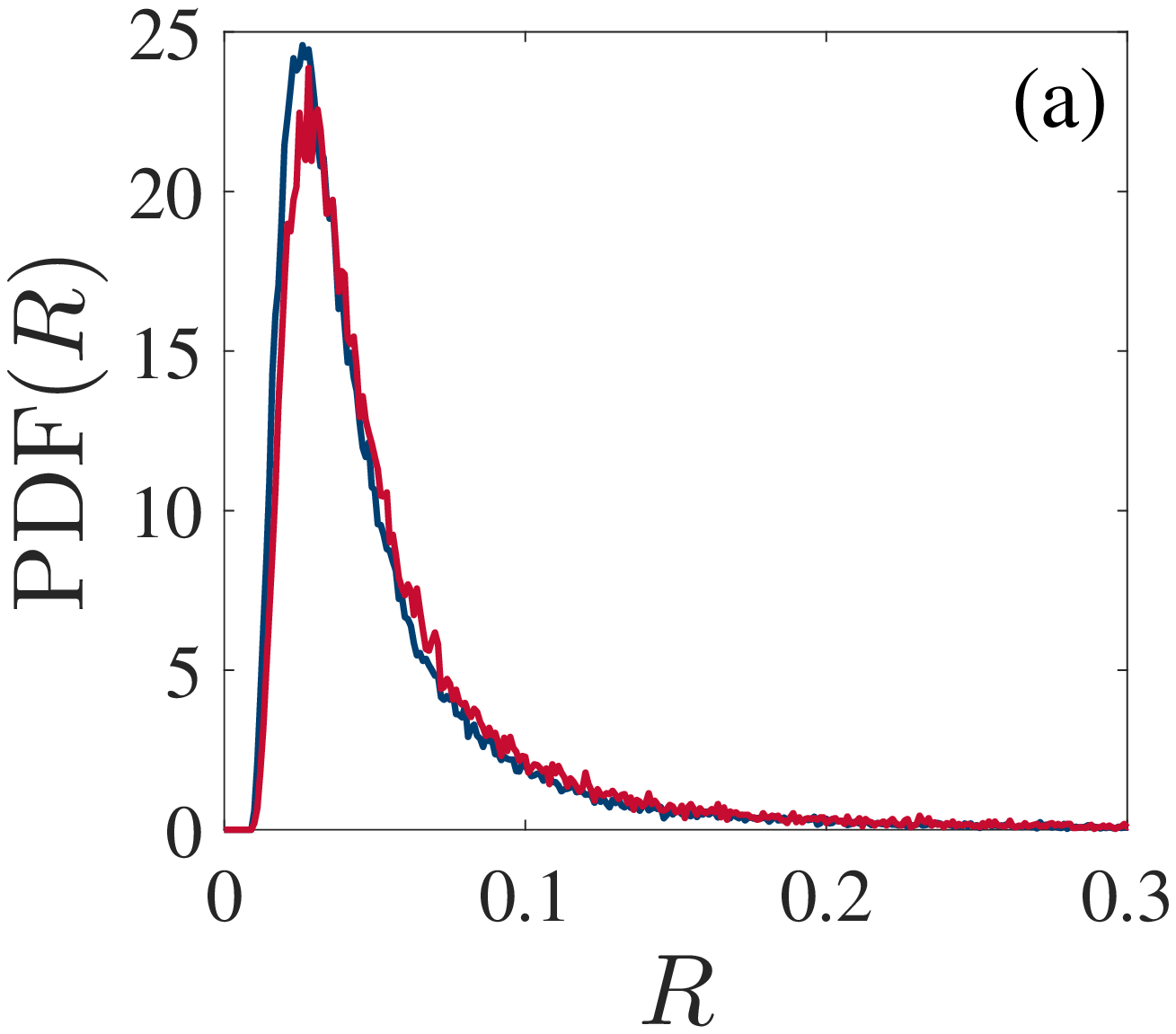}\includegraphics[width=0.5\linewidth]{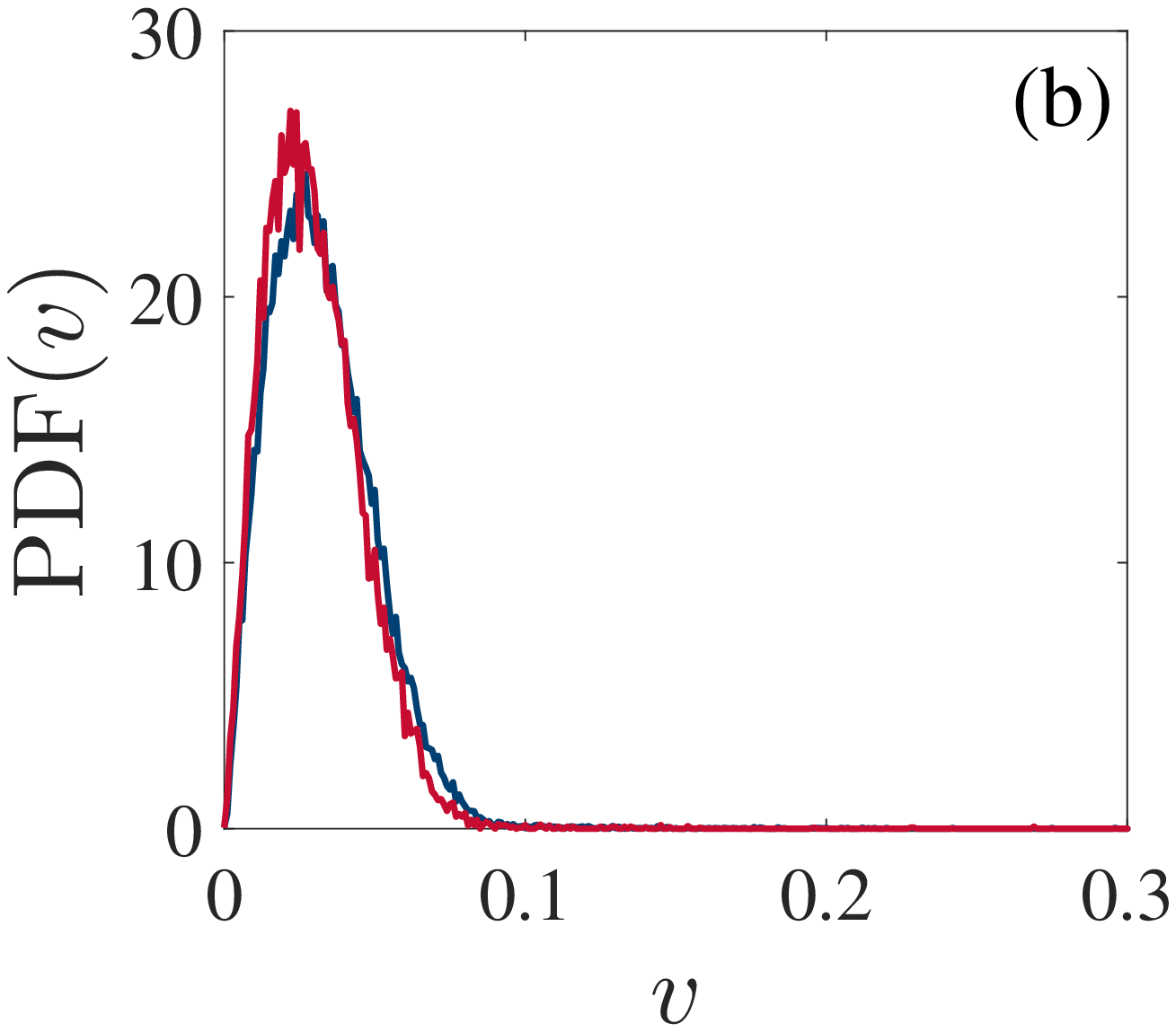}
\caption{Probability density functions (PDF) of the radius of curvature $R$ (a) and velocity $v$ (b) of vortex filaments retained after disregarding loops containing fewer than $200$ points. Red lines correspond to the lower density simulations, blues lines to the higher density simulations.} 
\label{fig:curv}
\end{figure}

We note that most of these large loops forming the vortex cluster and used in our analysis are not circular vortex rings, but complex, often knotted vortex structures. Two such structures taken from two realisations are shown in Fig. \ref{fig:knots} for illustration purposes. The two objects account for more than 54\% (a) and more than 67\% (b) of the total linelength, and more than 80\% (a) and more than 87\% (b) of the total cluster linelength retained in our analysis. The distributions of the local radius of curvature and local velocity of the loops retained in our analysis are shown in Fig. \ref{fig:curv}. Note that the peak radius of curvature, at around $0.025$cm, is considerably smaller than the radius of a circular ring consisting of 200 points at our chosen resolution ($\sim0.64$cm), and the peak velocity is correspondingly higher. The local behaviour of these structures is like that of far smaller vortex rings. Furthermore, if the collective behaviour were ballistic, as for a system of isolated loops, we could expect vortices to drift by around $25$cm by $10^3$s, an order of magnitude greater than the deviation seen in Fig. \ref{drms}.

Note that if we remove the evaporating loops from the 
analysis based on the modified Vinen equation (which the authors
of Ref.~\cite{Tsubota-2003}
did not do), our estimate for $\nu^{\prime}$ is (perhaps unsurprisingly) reduced. 
Indeed, without the evaporating loops, we obtain
$\nu^{\prime}/\kappa=0.19\pm0.08$ in the higher density simulations, 
and $\nu^{\prime}/\kappa=0.08\pm0.08$ in the lower density ones.


\section{Discussion}

In conclusion, we have found that a cluster of turbulent vortex lines,
initially localised in a region of space, spreads out
driven by two
effects: the evaporation of small vortex loops which leave the
cluster, and the slower spread of the cluster itself. The latter effect
can be modelled as a diffusion process which apparently
emerges in this inviscid 
fluid context from the interaction between the vortex lines. 
Using the standard approach based on rms deviations, we have found that 
the effective diffusion coefficient, 
measured in units of the quantum of circulation $\kappa$, 
is $\nu'/\kappa \approx 0.5$, independently of the initial 
vortex line density. Our finding
agrees quantitatively with values in the range $0.3 < \nu'/ \kappa < 0.5$
obtained in a 2D trapped atomic Bose-Einstein condensate using the GPE
model \cite{Rickinson-2018}, keeping in mind that in these confined systems 
$\nu'/\kappa$ seems to be reduced by 
boundary effects (vortex images).
It must be stressed that in both 2D and 3D, 
when determining the effective diffusion coefficient using the rms technique, 
we do not include the evaporating vortex loops 
because, unlike the vortices in the main cluster
which undergo continual collisions, 
they move freely at constant speed.

Our results contrasts with the smaller value reported by
Tsubota \etal \cite{Tsubota-2003}, $\nu'/\kappa=0.1$,  
which was obtained by fitting the solution of
a modified Vinen equation.
If we analyse our data with the modified Vinen equation
we obtain a similar lower estimate for $\nu'$ but with significant
error bars and sensitivity to the initial vortex line density;
another drawback of this technique is that
it requires independent knowledge of Vinen's parameter $\chi_2$ as described in the Appendix. On the contrary, the technique based on the rms deviations of tracers' trajectories
determines $\nu'$ more accurately and consistently, and gives values
comparable to previous findings in 2D.

It is also worth commenting on the difference between our initial conditions, 
and the initial condition used in \cite{Tsubota-2003}. In \cite{Tsubota-2003}
the vortex tangle was generated by a thermal counterflow
using the local induction approximation. This approximation in itself 
is known to be problematic in the presence of counterflow \cite{Adachi2010};
moreover a counterflow tangle is well known to be 
slightly anisotropic \cite{Adachi2010}. It seems plausible that 
this initial anisotropy in the initial condition used in \cite{Tsubota-2003}
modifies the diffusion of the tangle. 
Indeed, a study on the effect of anisotropy on the diffusion of quantum 
vorticity could prove fruitful.

The theory of Nemirovskii \cite{Nemirovskii-2010} yields a value
four times larger than ours, $\nu'/\kappa = 2.2$. 
This theory, as the modified Vinen equation,
assumes (rather than infers) the existence of
a diffusion process of vortex loops by postulating Brownian character
of the vortex loops' dynamics. A superfluid's effective viscosity is also
discussed in the different but related problem of the decay of superfluid
turbulence, with experimental and numerical
values \cite{Walmsley-2014,Babuin-2014,Zmeev-2015,Gao-2018} 
approximately in the range $0.01 < \nu'/\kappa < 1$ but more
concentrated around $0.1$.

Finally, we remark that the rms deviation method which we have used
to estimate $\nu^{\prime}/\kappa$, being Lagrangian, could be used
in future experimental studies of superfluid turbulence
using the newly developed
visualisation techniques based on excited helium molecules
\cite{mckinsey2005,rellergert2008,guo2009}.

\begin{acknowledgements}
CFB acknowledges the support of EPSRC grant
EP/R005192/1.
\end{acknowledgements}

\section*{Appendix: Determining $\chi_2$}
\begin{figure}[h]
\includegraphics[width=01.0\linewidth]{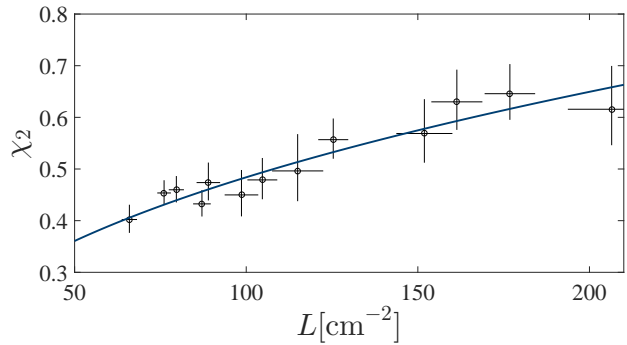}
\caption{Line density vs. $\chi_2$ calculated from steady state simulations (black circles) with errorbars showing $1$ standard deviation, with fit in blue. Inset: Steady state line density as a function of the injection rate, with fit in blue.}
\label{fig:Chi2}
\end{figure}

To estimate Vinen's parameter $\chi_2$ at zero temperature, and its dependence on the vortex line density $L$, we perform numerical simulations where identical vortex loops are continuously injected into a periodic box at random positions and orientations, at a constant rate. This gives us a known rate, $L_{\text{Inj}}$, at which vortex line length is injected. The simulation is continued until a saturated value of $L$ is achieved. When we are in this regime, the usual Vinen equation \cite{geurst1989}, which is usually written as:

\begin{equation}
\frac{\partial L}{\partial t} = \chi_1B\rho_n\vert v_{ns}\vert L^{3/2}-\frac{\kappa}{2\pi}\chi_2L^2,
\end{equation}
where $\rho_n$ is the normal fluid density, $v_{ns}$ the counterflow velocity, $B$ is a mutual friction coefficient, $\chi_1$ and $\chi_2$ two dimensionless parameters, reduces to:

\begin{equation}
0=L_{\text{Inj}}-\frac{\kappa}{2\pi}\chi_2L^2,
\end{equation}
as we have a (statistically) steady value for the line density, and our loop injection replaces the usual finite temperature source term from the normal fluid interacting with the vortex lines. This immediately gives $\chi_2=\frac{2\pi L_{Inj}}{\kappa L^2}$. Repeating this procedure for a range of injection rates allows us to construct the plots shown in Fig. \ref{fig:Chi2}, from which we fit $\chi_2$ as a function of L, finding $\chi_2\approx0.07L^{0.4}$

In a recent paper \cite{gao2018} a similar method is employed to estimate value of $\chi_2$ at finite temperatures, using thermal counterflow as a source term. Values for $\chi_2$ at $1.4$K (the lowest temperature reported) are reported as $2.10\pm0.34$ for $L=(3.59\pm0.34)\times10^3\text{cm}^{-2}$, $2.04\pm0.19$ for $L=(6.54\pm0.30)\times10^3\text{cm}^{-2}$, and $1.97\pm0.13$ for $L=(10.00\pm0.27)\times10^3\text{cm}^{-2}$. From our simulations we estimate $\chi_2=2.17$ for $L=3.59\times10^3$, $\chi_2=2.79$ for $L=6.54\times10^3$, $\chi_2=3.33$ for $L=10.00\times10^3$. Our values are consistent to order of magnitude, although slightly higher, possibly due to dissipation arising from the numerics. We note that for small $L$, dissipation is dominated by Kelvin wave stimulated emission of phonons, rather than by reconnections, so the assumption that dissipation scales as $L^2$ breaks down at some point, and accordingly our estimate of $\chi_2$ for small $L$ should be treated with caution.

~\\

\end{document}